\documentclass[usenatbib]{mn2e}

\usepackage[latin1]{inputenc}
\usepackage{amsmath}
\usepackage{amsfonts}
\usepackage{amssymb}

\usepackage{graphicx}
\usepackage{subfigure}
\usepackage{url}
\usepackage{mathrsfs}

\def \be{\begin{equation}}
\def \bea{\begin{eqnarray}}
\def \eea{\end{eqnarray}}
\def \ee{\end{equation}}
\def \Ms{M_{\odot}}
\def \a{\alpha}
\def \b{\beta}
\def \g{\gamma}
\def \h{\mathfrak{h}}
\def \A{{\cal A}}
\def \fr{\left( \frac{2}{9} \right)}
\def \hf{\frac{1}{2}}
\def \qtr{\frac{1}{4}}
\def \th {\tilde {h}}
\def \fl{f_{\rm lower}}
\def \fu{f_{\rm upper}}
\def \no {\nonumber}
\def\lsim{\mathrel{\rlap{\lower4pt\hbox{\hskip1pt$\sim$}}
    \raise1pt\hbox{$<$}}}                
\def\gsim{\mathrel{\rlap{\lower4pt\hbox{\hskip1pt$\sim$}}
    \raise1pt\hbox{$>$}}}                

\begin{document}

\title[Gravitational wave signature of a mini creation event (MCE)]{Gravitational wave signature of a mini creation event (MCE)}

\author[S.V. Dhurandhar and J.V. Narlikar]{ S.V. Dhurandhar and J.V. Narlikar\\ 
Inter-University Centre for Astronomy and Astrophysics, Post Bag 4, Ganeshkhind, Pune 411 007, India.}

\date{}
\maketitle

\begin{abstract} 
In light of the recent discoveries of binary black hole events by the LIGO detectors, we propose a new astrophysical source, namely, the mini creation event (MCE) as a possible source of gravitational waves (GW) to be detected by LIGO. The MCE is at the heart of the quasi steady state cosmology (QSSC) and is not expected to occur in standard cosmology. Generically, the MCE is anisotropic and we assume a Bianchi Tpye I model for its description. We compute its signature waveform and assume masses, distances analogous to the events detected by LIGO. By matched filtering the signal we find that, for a broad range of model parameters, the signal to noise ratio of the randomly oriented MCE is sufficiently high for a confident detection by advanced LIGO (aLIGO). We therefore propose the MCE as a viable astrophyical source of GW.          

\end{abstract}
\section{Introduction} 

The detection of gravitational waves in 2016 by LIGO has opened out a new avenue for the study of the cosmos.  The detection itself indicates that there are 
cosmic sources which cannot be detected by electromagnetic radiation but whose gravitational signature is open for detection.  Thus the first source to be 
detected by the LIGO detector consisted of coalescing black holes (Abbott et al. 2016a) neither of which could be detected by the conventional telescopes using electromagnetic 
radiation.  The two more sources subsequently detected have similar characteristics (Abbot et al. 2016b, 2017).  

Given that we now have a new method of detection, it is desirable that new sources which can be detected by the present LIGO techniques are looked for.  
This will make the existing detectors more versatile.  A similar development occured with detectors of electromagnetic waves when pulsars, quasars, X-ray
sources, gamma ray bursts etc. were detected.  Thus the possibility of a new astrophysical source will always be of interest to the LIGO type gravitational
wave detectors.  

The purpose of this paper is to propose a new source whose signature will be identified through match filtering methods used to extract signal from noise of a
gravitational wave detector.  The source proposed is the so-called mini-creation event (MCE) which was discussed in earlier papers (Das Gupta and Narlikar 1993,
Sarmah et al. 2006 and Narlikar et al. 2015)  The MCEs are pockets of matter creation distributed all over the universe, essentially replacing the big bang of 
standard cosmology.  The large scale behaviour of the universe thus is determined by the MCEs.  

We summarize in the following section the salient features of the universe vis-a-vis the MCEs, the new cosmology being known as the quasi-steady state 
cosmology (QSSC in brief).  

\section{The minicreation events (MCEs)} 

\subsection{The basic mechanism}

The details of the QSSC may be found in a series of papers by its authors Fred Hoyle, Geoffrey Burbidge and J.V. Narlikar (1993, 1994a, 1994b).  The overall 
background to the work has been discussed in a book by the same authors (Hoyle, Burbidge and Narlikar 2000). Broad features of the theory relevant to the work of this paper may be summarized as follows.  

\noindent (i) The field equations from which the QSSC models are derived are those obtained by Hoyle and Narlikar (1964) from an action at a distance formulation 
of Mach's principle.  Also,  an important difference from standard relativity is that the cosmological constant is negative.  \\

\noindent (ii)  Like the steady state theory of Bondi \& Gold (1948) and Hoyle (1948), this theory gives nonsingular models with matter creation sustained by a
negative energy scalar field $C$.  \\

\noindent (iii) The $C$-field produces matter at a typical spacetime point provided the particles comprising of the created matter satisfy the equality 

\be
m^2 = C_iC^i
\ee 

\noindent where $m$ is the mass of created particle and $C_i$ stands for $\partial C / \partial x^i$, $x^i (i=0,1,2,3,)$ being the time space coordinates. Theory
suggests that the created matter is in the form of Planck particles of mass $(\hbar c/G)^{1/2} \equiv m_P.$  \\

\noindent (iv)  In general this condition is not satisfied since the density of the $C$-field, $C_iC^i$, is very low.  But near a collapsed object of mass $M$, the 
value of $C_iC^i$ is raised.  If $R$ is the radius of the collapsed object, then $C_iC^i$ is raised by a factor 

\be
\gamma = \bigg(1 - \frac{2GM}{c^2R}\bigg)^{-1/2}.
\ee 

Thus the creation condition is satisfied provided $R$ is small enough and near $2GM/c^2$.  \\

\noindent (v)  A collapsing massive object in this theory is not headed for a black hole state followed by singularity but bounces at a finite radius $R_{{\rm min}}$ because of
the repulsive effect of the $C$-field owing to its negative energy.  Thus provided 

\be
m^2 \bigg(1 - \frac{2GM}{c^2R_{\rm min}}\bigg)^{-1/2} > m^2_P \equiv \frac{\hbar c}{G}
\ee 

\noindent we have creation of matter near highly compact massive objects. 

Here we assume that there are some massive objects which raise the level of the $C$-field such that the above inequality is satisfied for typical particle mass $m$.   Such massive objects act as centres of explosive creation because any new particle created is pushed outwards by the negative energy $C$-field owing to the repulsive effect produced by its negative energy. \\   

\noindent (vi)  Thus instead of a massive object becoming a black hole, it becomes a centre of creation.  This is called a minicreation event (MCE).  We may look upon 
any cosmic explosion as an MCE.  An MCE may also be referred to as a `minibang'.  Although its outward behaviour may be like that of big bang it differs from the latter
because it does not have a singular beginning, and its explosive creation does not violate the fundamental law of matter-energy conservation.  \\

\noindent (vii)  The universe as a whole responds to these MCEs happening all over it.  It can be shown that the universe has a long term steady expansion at an exponential 
rate along with a short term behaviour of oscillatory nature.  A simplified scale function of the universe is approximated by 

\be
S = {\rm exp}(t/P)\{1 + \xi {\rm ~cos~}(2 \pi t/Q)\}.
\ee 

The short-term oscillations reflect the condition whereby the MCE population pulsates up and down.  At the minimum of $S(t)$ the $C$-field is strong and more massive
objects satisfy the inequality (3).  This results in local expansion increasing.  As $S(t)$ increases $C^iC_i$ decreases and the massive objects drop out of the creative mode.  This 
ultimately gives rise to a contraction of the universe because of negative $\lambda$.  Also, because $| \xi |$ is less than unity $S(t)$ never becomes zero.   Typically, 
we may take $P \approx 10^3$ Gyr and $Q \approx 50$ Gyr. Also, we take $\xi = 0.8$ to fix ideas. \\

\noindent (viii)  The QSSC does not have a singularity, nor does it have a high temperature phase.  The maximum redshift observed in the cosmology is not expected to go 
beyond the range of 10-20.  Nevertheless the cosmology explains the observed microwave background radiation (MBR) as well as the abundances of light nuclei, vide
reference (Hoyle et al. 1994b) for MBR and (Hoyle et al. 1993) for nucleosynthesis.  The newly created Planck particle decays in a time scale of $\sim 10^{-43}$ sec.  The decay of Planck particle into 
baryons, leptons, etc. leads to local high temperature.  The `Planck fireball' evolves and leads to the observed light nuclei (Hoyle, et al. 1993).  The oscillatory cycle which 
lasts for $\sim 50$Gyr is sufficient for ordinary (Sun-like) stars to be born, evolve and then decay or explode leaving behind dark remnants.  It is suggested that these remnants contribute the observed dark matter.  What happens to the light emitted by stars in a cycle?  This is thermalized 
and is seen as MBR.  

This is a brief survey of the QSSC to which we now add gravitational waves as providing additional checks.  As was shown in the previous paper (Narlikar et al. 2015) that apart from discrete
source observations studies of continuum gravitational wave background can in principle be
compared with the post-inflation background produced in standard cosmology.  Thus, here is a test of cosmological background produced by sources.  Such a test may be possible at a future date.  

While a comparison of backgrounds of gravitational waves is a possible way of distinguishing between different cosmological models, a more practical method of
checking cosmological predictions is to try and detect an MCE.  For an MCE is required by the QSSC whereas it is not expected to occur in standard cosmology.  The recent detections of  discrete gravitational wave emitters mentioned before of black hole binaries, however, suggest that at the current level of detection technology, looking for specific sources is likely to be a more fruitful approach.  

Gravitational wave events were observed by the advanced LIGO detectors in their first run O1 and currently the second run is in progress. The second run has already made one confirmed detection (Abbot et al. 2017) and expects to observe many more events. So far the events observed are those of mergers of binary black holes whose signature waveforms have been computed analytically and numerically. Therefore, one knows what one is looking for and uses the match filtering methods (Sathyaprakash and Dhurandhar 1991) to extract the signals from the noise. It is possible that the data contain signals from other astrophysical sources and if so one should endeavour to detect and identify such signals. The astrophysical source we propose here is the MCE. However, in order to detect such a source one needs to know the signature of the signal by computing the waveform. In this calculation we compute the GW waveform based on the model of the MCE proposed by Narlikar and DasGupta (1993).   

\subsection{Gravitational radiation from an arbitrarily oriented anisotropic MCE}

Although a typical MCE is nonsingular in origin, being made of ejected newly created matter, the model assumed here will be approximated by a triaxial ellipsoid expanding 
anisotropically in all directions.  While in general relativity such a solution is singular, in our modified theory it will have arisen by a bounce at a small size.  Thus replacing this minimum size
by zero will not produce much error.  
 
Because of the anisotropy we expect the MCE to emit gravitational waves. The expansion is described via a Bianchi Type I model whose metric is given by:
\be
ds^2 = c^2 dt^2 - X^2 (t)~dx^2 - Y^2 (t)~dy^2 - Z^2 (t)~dz^2 \,,
\ee
where $(x,y,z)$ are the comoving coordinates and $t$ is the proper time of a dust particle moving outwards. Solving Einstein's equations for dust dominated systems we have:
\bea
X(t) &=& S(t) [F(t)]^{2 \sin \g} \,, \no \\
Y(t) &=& S(t) [F(t)]^{2 \sin (\g + 2 \pi/3)} \,, \no \\
Z(t) &=& S(t) [F(t)]^{2 \sin (\g + 4 \pi/3)} \,,
\eea
where,
\be
F(t) = \frac{(GM)^{1/3} t^{2/3}}{S (t)},~~~~~~~S^3 (t) = X(t) Y(t) Z(t) \,.
\ee
The anisotropy is related to the parameter $\g$ which varies between $- \pi/6$ to $\pi/2$. The average scale factor $S$ is related to the mass $M$ by:
\be
S^3 (t) = \frac{9}{2} GMt(t + \Sigma), ~~~~~~~\Sigma = {\rm const.} \,.
\ee  
Such an expanding object has time varying quadrupole moment and will emit gravitational waves.  Again, we stress that the presence of singularity at $t=0$ will not alter the conclusion in any significant way.  
\par
 
The source frame of the MCE is denoted by $(x, y, z)$ and the MCE has principal axes as the coordinate axes. Thus the quadrupole tensor is diagonal because of the inherent symmetry assumed in the model and can be easily computed - there are no off-diagonal terms. It is given by:
\be
I_{xx} = \frac{1}{5} M X^2 (t),~I_{yy} = \frac{1}{5} M Y^2 (t),~I_{zz} = \frac{1}{5} M Z^2 (t) \,.
\label{inertia}
\ee
We consider the situation when $t >> \Sigma$. Then the following simplifications occur. We have:
\be
S (t) = \left (\frac{9}{2} GM \right )^{1/3} t^{2/3}, ~~~~~~~F (t) = \left (\frac{2}{9} \right )^{1/3} \,.
\ee
Also,
\be
X(t) = (GM)^{1/3} t^{2/3} \left (\frac{2}{9} \right)^{\frac{2}{3} \sin \g - \frac{1}{3}} \,,
\label{metric}
\ee
with $Y(t)$ and $Z(t)$ described by similar expressions where $\g$ is replaced by $\g + 2 \pi/3$ and $\g + 4 \pi/3$ respectively. Note we could have switched to labelling the $(x, y, z)$ axes as $(x_1, x_2, x_3)$ but we do not do this in anticipation of what follows. We need to transform from the source frame to the wave frame in order to get the wave amplitudes and the usual convention in the literature is to use $(x, y, z)$ for the source frame and $(X, Y, Z)$ for the wave or radiation frame and so we follow this notation (Dhurandhar and Tinto 1988). 
\par
From these expressions the quadrupole tensor can be readily computed. From Eqs. (9) and (11),  its components are given by:   
\be
I_{xx} = \frac{1}{5} M (GM)^{2/3} t^{4/3} \left (\frac{2}{9} \right)^{\frac{4}{3} \sin \g - \frac{2}{3}} \,,
\ee
with similar expressions for $I_{yy}$ and $I_{zz}$ in which $\g$ is replaced by $\g + 2 \pi/3$ and $\g + 4 \pi/3$ respectively. 
\par
The GW strain amplitudes are proportional to the second time derivatives of the quadrupole tensor evaluated at the retarded time $t - R/c$, where $R$ is the distance from the observer to the MCE. Explicitly, 
\be
h^{TT}_{ik} (R, t) = \frac{2 G}{c^4} \frac{1}{R} \left [{\ddot I}_{ik} (t - R/c) \right]^{TT} \,.
\ee   
The superscript $TT$ refers to the transverse-traceless gauge. With this as our goal we define the basic GW strain amplitudes $\h_1, \h_2, \h_3$ which incorporate the second time derivative of the quadrupole tensor at the retarded time. These strain amplitudes appear in the final GW amplitudes in the radiation frame. We define:
\be
\h_k (\g) = \frac{\A}{R} \left (t - \frac{R}{c} \right )^{-2/3} \left (\frac{2}{9} \right )^{\frac{4}{3} \sin \left(\g + (k - 1) \frac{2 \pi}{3} \right)} \,, 
\label{basic}
\ee
where $k = 1, 2, 3$ and,
\be
\A = \frac{4}{5} \fr^{1/3} \frac{(GM)^{5/3}}{c^4} \,,
\label{amp}
\ee
is a constant amplitude. Note that these basic strain amplitudes $\h_k$ depend on the parameter $\g$.  
 
We next compute the two polarisation GW strain amplitudes in the wave or radiation frame. The source frame is denoted by the $(x, y, z)$ frame. The coordinate axes are also chosen to be the principal axes of the MCE. However, in general, the MCE can have arbitrary orientation and therefore the source frame can be arbitrarily oriented with respect to the observer. We therefore need to compute the two GW polarisation amplitudes denoted by $h_+$ and $h_{\times}$ in the wave frame. The wave frame is denoted by $(X, Y, Z)$. We therefore rotate the source frame to the wave frame by the Euler angles $\a, \iota, \b$ using the Goldstein convention - first rotation by angle $\a$ about the $z$-axis, then second rotation about the line of nodes (new x-axis) by angle $\iota$ and the final rotation about the new $z$-axis by angle $\b$. However, if we just need to point the new $z$-axis along the line of sight (negative $Z$ axis), then only the first two rotations are necessary, namely, by angles $\a$ and $\iota$. But then the orientation of the $X-Y$ axes will be determined by the orientation of the source - this may be sufficient for certain purposes, but in general is not desirable. If we fix the wave frame then another rotation by the angle $\b$ is necessary to make the rotated $(x, y)$ axes coincide with the $(X, Y)$ axes. Here we give the GW amplitudes for both situations.
\par
The transverse and traceless components of the metric perturbation in the wave frame give the two GW polarisation amplitudes. We first consider only the two rotations by angles $\a$ and $\iota$. We call these amplitudes $h_{+ 0}$ and $h_{\times 0}$. They are given by:
\bea
h_{+ 0} &=& \hf (1 + \cos^2 \iota)~ \cos 2 \a ~ h_1 (\g) ~+ ~ \sin^2 \iota ~ h_2 (\g) \, \no \\
h_{\times 0} &=& \cos \iota ~\sin 2 \a ~ h_1 (\g) \,,
\eea
where,
\be
h_1 (\g) = \hf (\h_1 - \h_2) \,,~~~ h_2 (\g) = \qtr (\h_1 + \h_2 - 2 \h_3) \,,
\ee
where $\h_1, \h_2, \h_3$ have already been defined in Eq. (14).
\par
Including also the third rotation by the angle $\b$, we can write the final GW strain amplitudes $h_+$ and 
$h_{\times}$ in the wave frame in terms of the amplitudes $h_{+ 0}$ and $h_{\times 0}$ just by using the tensor transformation law:
\bea
h_+ &=& h_{+ 0} ~\cos 2 \b ~ - ~h_{\times 0}~\sin 2 \b \,, \no \\
h_{\times} &=& h_{+ 0} ~\sin 2 \b ~ + ~h_{\times 0}~\cos 2 \b \,.
\eea
We can now explicitly write these amplitudes in terms of $h_1 (\g)$ and $h_2 (\g)$. 
\bea
h_+ &=& \left [\hf (1 + \cos^2 \iota) \cos 2 \a \cos 2 \b - \cos \iota \sin 2 \a \sin 2 \b \right ]h_1 (\g) \, \no \\
    && ~+~ \sin^2 \iota ~\cos 2 \b~ h_2 (\g) \, \no \\
h_{\times} &=& \left [\hf (1 + \cos^2 \iota) \cos 2 \a \sin 2 \b + \cos \iota ~\sin 2 \a \cos 2 \b \right ]h_1 (\g) \, \no \\
    && ~+~ \sin^2 \iota ~\sin 2 \b~ h_2 (\g) \,.
\eea

\section{Astrophysical MCE and their detection by LIGO}

In order to fix ideas, let us first consider the simple case of an MCE whose $z$ axis points along the line of sight and also the $(x, y)$ axes coincide with the $(X, Y)$ wave axis so that $\a = \b = \iota = 0$. This implies that only $h_+ \neq 0$, that is, $h_\times = 0$ and that $h_+ = h_1 (\g)$. We therefore take the GW strain $h_{xy}$ to be just $h = h_+ \equiv h_1 (\g)$. We then find,
\bea
h_{xy} &=& \hf (\h_1 - \h_2) \, \no \\
  &=& \frac{\A}{R} ~\tau^{-2/3} ~\eta_{xy} \,,
\eea
where the amplitude $\A$ has been defined in Eq. (\ref{amp}),  $\tau = t - R/c$ is the retarded time and the anisotropy parameter $\eta_{xy}$ is given by,
\be
\eta_{xy} (\g) = \hf \left[ \fr^{4/3 \sin \g} - \fr^{4/3 \sin (\g + \frac{2 \pi}{3})} \right] \,.
\ee 
Note that the $\eta_{xy}$ defined here is half that of defined in Das Gupta and Narlikar 1993. The anisotropy parameter takes values of the order of unity over the range of permissible $\gamma$. A plot of $\eta_{xy}$ versus $\gamma$ has been shown in Figure 1 as the dashed curve. We now evaluate $h_{xy}$ for typical astrophysical parameters. We take these parameters to be about those of the recent black hole binary events discovered by LIGO. For example, the last event discovered GW170104 in O2 has a mass of about $50~ \Ms$ and was at an estimated luminosity distance of 880 Mpc. We therefore take the mass of the typical MCE to be $\sim 50~ \Ms$ and to be at a distance of $\sim 1$ Gpc. These values give,
\be
h_{xy} \sim 4.57 \times 10^{-24} \left (\frac{M}{50~\Ms} \right)^{5/3} \left (\frac{R}{\rm Gpc} \right)^{-1} \tau^{-2/3} \eta_{xy}\,.
\ee   
This is well within the range of the LIGO detectors. However, this is not the largest value that the GW strain can achieve for the MCE considered. In fact we find that this MCE model is more anisotropic in the $(x, z)$ and $(y, z)$ axes and thus will produce larger GW amplitudes orthogonal to these axes. In fact, the fair course to take is to average the GW strain over all orientations. Since we have no apriori knowledge, we will assume a uniform distribution of orientations and average accordingly over the angles $\a, \b$ and $\iota$. The result is that we obtain an average or root mean square (rms) value of the GW strain which we denote by $h_{\rm rms}$ which in turn can be expressed in terms of the anisotropy parameter $\eta_{\rm rms} (\g)$. We then have:
\be
\eta_{\rm rms} (\g) = \frac{2}{\sqrt{15}}[\eta_{xy}^2 + \eta_{yz}^2 + \eta_{zx}^2 ]^\hf \,.
\ee
In Fig. 1 the dashed curve represents $\eta_{xy} (\g)$ and the continuous curve represents $\eta_{\rm rms} (\g)$. We show the plot below:
\begin{center}
\begin{figure}
{\includegraphics[width=0.4\textwidth]{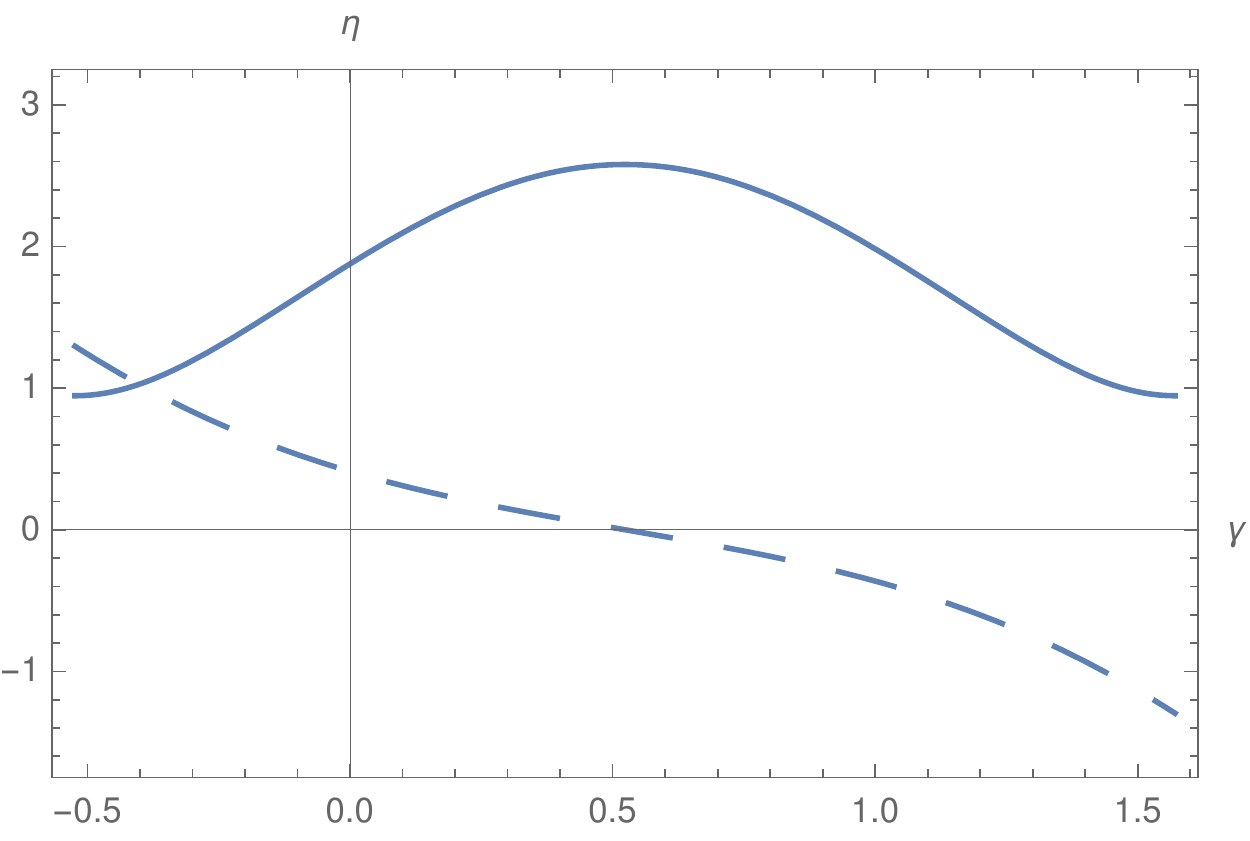}}
\caption{The dashed curve shows $\eta_{xy}$ corresponding to the $(x, y)$ axes and the continuous curve shows $\eta_{\rm rms}$ uniformly averaged over all orientations as a function of $\g$ where $-\pi/6 \leq \g \leq \pi/2$.}
\label{fig1}
\end{figure}
\end{center}
In terms of $\eta_{\rm rms}$ the GW strain is given by,
\bea
h_{\rm rms} (\tau) &=& \frac{\A}{R}~ \eta_{\rm rms}~ \tau^{-2/3} \,, \no \\ 
&\simeq& 4.57 \times 10^{-24} \left (\frac{M}{50~\Ms} \right)^{5/3} \left (\frac{R}{\rm Gpc} \right)^{-1}~\eta_{\rm rms} \no \\
&\times& \tau^{-2/3} \,.
\label{h_timedom}
\eea 
From Figure 1, we see that $\eta_{\rm rms}$ can go upto almost $2.6$ and therefore such a source should be observable by the advanced LIGO detectors. In order to check this we must compute the signal to noise ratio (SNR) of such an arbitrarily oriented event. For this we need the GW strain $h_{\rm rms}$ in the Fourier domain. Taking the Fourier transform of $h_{\rm rms}(\tau)$ from Eq.(\ref{h_timedom}) and then taking its absolute value for $f > 0$ (only this is needed to compute the SNR), we obtain:
\be
|\th_{\rm rms}(f)| \simeq 6.63 \times 10^{-24} \left (\frac{M}{50 \Ms} \right)^{5/3} \left (\frac{R}{\rm Gpc} \right)^{-1} \eta_{\rm rms} f^{-1/3} \,.
\ee
We now go on to calculate the SNR. We use the one sided noise power spectral density (PSD) of aLIGO corresponding to zero detuned high power. This is normally given numerically, but for our purpose the analytic fit given by Ajith P. (2011) suffices. The results will at most differ by a percent or so which is acceptable under the circumstances. The analytical fit to the noise PSD is given by:
\bea
S_h (f) &=& 10^{-48}(0.0152 x^{-4} + 0.2935 x^{9/4} + 2.7951 x^{3/2} \, \no \\ 
&& - 6.5080 x^{3/4} + 17.7622) \,,
\eea
where $x = f/245.4$. The lower frequency cut-off is assumed to be 20 Hz.  The SNR which we denote by $\rho$ of the rms signal is then given by:
\be
\rho = 2 \left [\int_{\fl}^{\fu} df ~\frac{|\th_{\rm rms} (f)|^2}{S_h (f)} \right]^\hf \,.
\ee
Taking the lower cut-off $\fl$ to be 20 Hz and the upper cut-off $\fu$ to be 1 kHz, we then obtain:
\be
\rho (\g) \sim 14.36 \times \left (\frac{M}{50~\Ms} \right)^{5/3} \left (\frac{R}{\rm Gpc} \right)^{-1} \eta_{\rm rms} (\g) \,.
\ee
In this frequency band most of the SNR is accumulated and yields fairly accurate results. Since a typical value of $\eta_{\rm rms} \sim 2$, we typically have $\rho \sim 30$. When $\g \sim \pi/6$, $\eta_{\rm rms}$ attains more or less its maximum value $\eta_{\rm rms} \sim 2.58$ for which we have $\rho \sim 37$.
\par

However, the SNR we have calculated is for a detector orientation most favourable to the source direction. But in general this will not be the case and the SNR will be reduced because of angular factors. For a source direction described by the angles $\theta, \phi$ in the frame of the detector, the signal $h(t)$ is given by:
\be
h(t) = h_+ (t) F_+ (\theta, \phi) + h_{\times} F_{\times}(\theta, \phi) \,,
\ee
where $F_+ (\theta, \phi), F_{\times} (\theta, \phi)$ are the antenna pattern functions (see Dhurandhar and Tinto 1988) given in terms of the direction angles $\theta, \phi$. Since we do not know from which direction the signal will arrive, and given our lack of any apriori knowledge, we may assume a uniform distribution of sources over the sky directions and then average over the sky directions. A simple calculation shows that the reduction factor is $2/5$ in the signal amplitude and therefore also in the SNR. For the typical value of $\eta_{\rm rms} \sim 2$, the sky averaged SNR will be $\sim 12$ and for maximum value of $\eta_{\rm rms}$, the sky averaged SNR $\sim 15$. Such high SNRs for black hole binaries generally imply confident detection. We may expect this to be the case here also. However, a detailed analysis, which we do not perform here, involving real data is necessary in order to confirm this.

\section{Future Outlook}

In this article, we make a case for the MCE as a possible astrophysical source for GW astronomy. The recent discoveries of the binary black hole events were made with the two LIGO detectors situated at Hanford and  Livingston, U. S. In the current situation of the data containing non-Gaussian, non-stationary noise, more than a single detector with uncorrelated noise is required to make a detection with acceptable confidence. Moreover, using more than one detector increases the SNR; for two detectors of similar sensitivity operating simultaneously, the SNR will increase by a factor of $\sqrt{2}$. In the estimates of the SNR that we have obtained here, we have assumed a single LIGO detector operating at design sensitivity. If one were to consider making a detection of the MCE from current detector data or the data that we expect in the near future, procedure analogous to the one for detecting binary black holes will have to be followed. One would have to first match filter the data, look for coincidences in the time of arrival and most importantly estimate the noise background. The noise background can be estimated by performing time slides, where one time slides the filtered output of one detector relative to the other with durations comfortably larger than the GW travel time between the detectors - this is $\sim 10$ milliseconds for the LIGO detectors. We would like to mention here that, in case of the MCE, the analysis would be simplified in one important way as compared to the one used in the detected binary black hole events - there is only one template for the simple model of the MCE that we have considered here - the signal has just one amplitude parameter, as compared to the hundreds of thousands of templates required for binary black hole coalescences where the waveforms depend on several parameters. Further, if more detectors come on line, the Italian-French Virgo detector, for example, or the Japanese KAGRA in few years time, or Ligo-India then the chances of detecting such events will greatly increase. GW astronomy is set to herald a new era in fundamental physics, cosmology and astrophysics.

\section*{References}

\noindent Abbott, B. P. et al, LIGO Scientific Collaboration, Virgo Collaboration, 2016a, Phys. Rev. Lett., 116, 061102 \\

\noindent Abbott, B. P. et al, LIGO Scientific Collaboration, Virgo Collaboration, 2016b, Phys. Rev. Lett., 116, 241103 \\

\noindent Abbott, B. P. et al, LIGO Scientific Collaboration, Virgo Collaboration, 2017, Phys. Rev. Lett., 118, 221101 \\

\noindent Ajith P., 2011, Phys. Rev. D., 84, 084037 \\

\noindent Bondi H., Gold T., 1948, MNRAS, 108, 252 \\

\noindent Das Gupta P., Narlikar J. V. , 1993, MNRAS, 264, 489  \\

\noindent Dhurandhar S. V., Tinto M., 1988, MNRAS, 234, 663 \\

\noindent Hoyle F., 1948, MNRAS, 108, 372 \\

\noindent Hoyle F., Narlikar J. V., 1964, Proc. Roy. Soc., A282, 191 \\

\noindent Hoyle F., Burbidge G., Narlikar J.V., 1993, ApJ, 410, 437 \\

\noindent Hoyle F., Burbidge G., Narlikar J.V., 1994a, MNRAS, 267, 1007 \\

\noindent Hoyle F., Burbidge G., Narlikar J.V., 1994b, A\&A, 289, 729 \\

\noindent Hoyle F., Burbidge G., Narlikar J.V., 2000, A Different Approach to Cosmology, 
Cambridge Univ. Press, Cambridge \\ 

\noindent Narlikar J. V., Dhurandhar S. V., Vishwakarma R. G., Valluri S. R., Auddy S., 2015, MNRAS, 451, 1390 \\ 

\noindent Sarmah B.P., Banerjee S.K., Dhurandhar S.V., Narlikar J.V., 2006, MNRAS, 369, 89 \\ 

\noindent Sathyaprakash B. S., Dhurandhar S. V., 1991, Phys. Rev. D, 44, 3819

\end{document}